\documentclass[conference]{IEEEtran}
\usepackage{cite}
\usepackage{amsmath}
\usepackage{algorithmic}
\usepackage{array}
\ifCLASSOPTIONcompsoc
  \usepackage[caption=false,font=normalsize,labelfont=sf,textfont=sf]{subfig}
\else
  \usepackage[caption=false,position=top,font=footnotesize,justification=raggedright,singlelinecheck=false]{subfig}
\fi
\usepackage{url}
\usepackage{float}
\usepackage{listings}
\floatstyle{ruled}
\newfloat{listing}{htbp}{lop}
\floatname{listing}{Listing}
\usepackage[pdftex]{graphicx}
\usepackage{xcolor,soul}
\usepackage{tikz}

\newcommand\copyrighttext{This work has been presented at the 27th International Conference on Modeling, Analysis and Simulation of Wireless and Mobile Systems (MSWiM 2025), Barcelona, Spain October 27th – 31st, 2025, and is included in the conference proceedings. Copyright is owned by IEEE.}
\newcommand\copyrightnotice{%
\begin{tikzpicture}[remember picture,overlay]
\node[anchor=south,yshift=20pt] at (current page.south) {\fbox{\parbox{\dimexpr\textwidth-\fboxsep-\fboxrule\relax}{\copyrighttext}}};
\end{tikzpicture}%
}
\makeatother
\begin{document}
\IEEEoverridecommandlockouts
\title{Toward Hybrid COTS-based LiFi/WiFi Networks with QoS Requirements in Mobile Environments}
\author{\IEEEauthorblockN{Emilio Ancillotti, Loreto Pescosolido, Andrea Passarella}
\IEEEauthorblockA{CNR Institute for Informatics and Telematics (IIT)\\
Via Giuseppe Moruzzi 1, Pisa, Italy\\
Email: \{emilio.ancillotti, loreto.pescosolido, andrea.passarella\}@iit.cnr.it }
}

\maketitle
\copyrightnotice
\begin{abstract}
We consider a hybrid LiFi/WiFi network consisting of commercially available equipment, for mobile scenarios, where WiFi backs up communications, through vertical handovers, in case of insufficient LiFi QoS. When QoS requirements in terms of goodput are defined, tools are needed to anticipate the vertical handover relative to what is possible with standard basic mechanisms, which are only based on a complete loss of connectivity. We introduce two such mechanisms, based on signal power level readings and CRC-based packet failure ratio, and evaluate their performance in terms of QoS-outage duration, considering as a benchmark an existing baseline solution based on the detection of a connectivity loss. In doing this, we provide insights into the interplay between such mechanisms and the LiFi protocol channel adaptation capabilities. Our experimental results are obtained using a lab-scale testbed equipped with a conveyor belt, which allows us to accurately replicate experiments with devices in motion. With the proposed methods, we achieve QoS outages below one second for a QoS level of 20~Mbps, compared to outage durations of a few seconds obtained with the baseline solution.\end{abstract}

\begin{IEEEkeywords}
Hybrid LiFi-WiFi, VLC, vertical handover.\vspace{-3.5mm}
\end{IEEEkeywords}

\section{Introduction}\vspace{-1.5mm}
Hybrid LiFi/WiFi networks \cite{Zeng2020,Wu2021} (see also references therein) have emerged as a promising wireless networking paradigm for many scenarios. Such networks are composed of small LiFi cells, named attocells, with a typical coverage radius of very few meters, e.g., 1.5~m, depending on the ceiling height, and one or more WiFi cells, which cover a wider region. LiFi Access Points (APs) use LED lamps as Visible Light Communication (VLC) transmitters, and infrared (IR) photodiodes as receivers. The LiFi interface of each mobile node is composed of an IR transmitter and a visible light photodiode receiver. In the simplest configuration, one of the two (LiFi and WiFi) network interfaces in each node is used as the default one, while the other has a backup role, i.e. it is used when the link provided by the default one is not available or starts performing poorly. The performance in the execution of vertical handovers, i.e., the process of diverting traffic from one interface to the other, is of paramount importance to avoid deterioration of the QoS.
Focusing on networks composed of commercial-off-the-shelf (COTS) devices, vertical handovers can be handled by mature software tools, e.g., for Linux-based systems, the Linux Ethernet Bonding Driver \cite{linuxbonding} (in the following, Bonding Driver, for short) or the Libteam library \cite{libteam}. Other solutions have been proposed in the literature, but are far from being practically usable in real-life network deployments, see Section~\ref{sec:Related-work}, or highly inefficient, as MTCP-based ones. The two above-mentioned tools have the advantage of being ready for use in a real-life network deployment, but have the disadvantage of triggering a handover only when the connectivity of a given interface is completely lost. They are not designed to deal with connectivity impairments which are not sufficiently severe to halt a packet flow on a given network interface.

Despite hybrid LiFi/WiFi networks have been studied for several years, most of the results in this area are based on analytical modeling and simulations, with experimental results limited to relatively few works covering selected specific aspects, see Section~\ref{sec:Related-work}. 

In this experimentally based study, we consider the problem of QoS decay on a packet flow for a mobile device in an indoor hybrid LiFi/WiFi network, composed of COTS, as the mobile device approaches the attocell border and eventually exits it. We assume that LiFi is the default interface in use, which should be replaced by WiFi as soon as the QoS drops below a prescribed level. What happens in a network properly configured with one of the above-mentioned tools, is that the connection sits on LiFi until it is almost completely lost (more details in  Section~\ref{sec:perf-eval}). This leads to large intervals during which, despite the LiFi connection being up, the link can be considered out of service from a QoS standpoint. Note that this happens independently from how fast the link monitoring tools are in triggering the vertical handover, because they do so only as the connection is lost, which could be way after the moment the QoS drops below a threshold defined in terms of goodput.

To deal with this problem, we propose using real-time signal power level and CRC-based packet failure ratio readings, which can be obtained by the network interface drivers using standard command tools. We first analyze, from a qualitative point of view, the features of the time evolution of these quantities, considering different speeds of the mobile device, and then propose two mechanisms based on setting suitable thresholds on the readings of these metrics to enforce vertical handovers. In this work, our experimental analysis is limited to the downlink case and the presence of traffic for a single user, although multiple active nodes are present in the experiment with their signalling traffic.

The paper is organized as follows: In Section~\ref{sec:Related-work} we review existing experimentally based works dealing with hybrid LiFi/WiFi networks, positioning our work in the research area. In Section~\ref{sec:network-model} we present the hybrid WiFi/LiFi network model and describe our testbed. In Section~\ref{sec:problem-statement} we highlight the limits of existing standard solutions, in scenarios where nodes move inside and outside an attocell, in terms of adaptability to QoS requirements, and propose two techniques to overcome these limits. Section~\ref{sec:perf-eval} presents our experimental results, showing the effectiveness of the proposed solutions in terms of QoS-outage duration. Finally, Section~\ref{sec:conclusion} summarizes our contribution.\vspace*{-4mm}

\section{Related work}\label{sec:Related-work}\vspace*{-1mm}
The majority of works on hybrid LiFi/WiFi networks are based on simulation and modeling results. In this section, we focus on works that include a testbed-based experimental contribution covering the integration of two communication technologies, which are the most relevant to our work.

In \cite{Zhang2017}, a proof-of-concept (POC) based on COTS (LEDs, IR transceivers, WiFi APs) is presented. Handovers are handled at the transport layer, using an MTCP-based implementation with no additional input from lower protocol stack layers. The solution covers downlink data transmission and requires the presence of a central coordinator which associates users to attocells and to a WiFi cell. Access information is periodically sent by the mobile devices, which exploit the beacon packets sent from the APs. The performance evaluation focuses on the average data rate of mobile devices moving across different attocells at a speed of 0.1 m/s, showing the gain obtained by the hybrid solution compared to a legacy single communication technology-based solution. The time granularity of the results does not provide information on time-localized connectivity losses, or temporary QoS decreases, around handover instants. In \cite{Saud2017}, an implementation using USRPs connected to PCs, visible light and infrared LEDs, photodiode receivers, and RF transceivers is described. The POC implements a point-to-point link. Link monitoring is implemented either by a custom protocol or by the Bonding Driver  \cite{linuxbonding}. In both cases, the device periodically sends control packets and, if on one link no reception is registered within a prescribed time threshold, the corresponding link (visible light or RF) is considered to be down. Neither \cite{Zhang2017} nor \cite{Saud2017} consider concurrent transmissions from multiple users.

Zeng \emph{et al.}, \cite{Zeng2020}, study dynamic load balancing schemes for a hybrid LiFi/WiFi network, considering performance metrics focused on the overall network capacity.  The integration of the two communication technologies is handled at layers above the TCP/IP one. With this approach, each physical network interface on the same device has a different IP address. Possibly, it may also belong to a different subnet.  Because a (quasi-) real-time tracking of the status of the physical links related to the different medium access technologies is necessary, this approach provides more flexibility, but adds the complexity of dealing with multiple IP addresses associated with the same device. The focus in \cite{Zeng2020} is on network capacity, whereas the main focus of our work is on the effect of handover at small timescales.

Integration of LiFi and WiFi at the lower layers of the protocol stack has also been proposed. In \cite{Zubow2021}, the integration is performed at the physical layer, directly on chip, by attaching VLC transducers to 802.11-compliant chips, to obtain a unique network interface. Although proving to be a promising solution, this approach has not yet led, to the best of our knowledge, to commercially available products, which are the target of our work. The performance evaluation in \cite{Zubow2021} focuses on the reliability and robustness of the proposed solution, interference handling, and overall data rate. From a performance point of view, a direct comparison of \cite{Zubow2021} with our work is not possible.

In \cite{Haas2020}, the authors investigate load balancing problems in Hybrid LiFi/WiFi networks by using, besides simulations, a real-life software-defined networking (SDN)-based testbed, with the aim to evaluate the increase in aggregate and per-user data rate with respect to a legacy WiFi solution. In this case, an application layer controller software runs in the APs, which manages both vertical (between LiFi attocells and WiFi cells) and horizontal (across LiFi attocells) handovers. The APs periodically report the status of the individual connections to the controller, which manages the nodes-APs associations. Overall, the solutions proposed in \cite{Haas2020} are proactive load balancing algorithms operating at time scales larger than those entailed by purely reactive vertical handovers as the Bonding Driver or Libteam. The performance evaluation is more focused on the average data rate, rather than on collecting the statistics on handover-induced outage intervals, although handover-induced outage durations on the order of 5 seconds are reported. This suggests that the link status is monitored by keeping track of the presence/absence of device-AP association as determined by the network interface. Our approach is, instead, that of a more aggressive link status monitoring, either assuming the use of a bonding mechanism between the interfaces, and specifically Libteam, or proposing solutions to anticipate the vertical handover, based on local signal power level or CRC readings.

Jarchlo \emph{et al.}, \cite{Jarchlo2022}, consider a hybrid LiFi/WiFi network, and propose a handover solution that operates at both the data link and transport layers. The data link component of the proposed solution uses the Bonding Driver. The proposed solution is completed by adding a multi-path transport control protocol (MPTCP)-based component at the transport layer. The performance in terms of vertical handover execution time is comparable to what reported in our previous work \cite{Pescosolido2022}, although the focus of \cite{Jarchlo2022} is not on this specific aspect and no mechanism is provided to increase the effectiveness of the handover in terms of adjustable QoS requirements.

Finally, to the best of our knowledge, this is the first experimentally based work, in the context of hybrid LiFi/WiFi networks, considering the Libteam library as a baseline link monitoring and interface switching tool as an alternative to the Bonding Driver.\vspace*{-1mm}

\section{Network Model and Testbed Description}\label{sec:network-model}\vspace*{-1mm}

\begin{figure}[t]
\begin{center}
\includegraphics[width=\columnwidth]{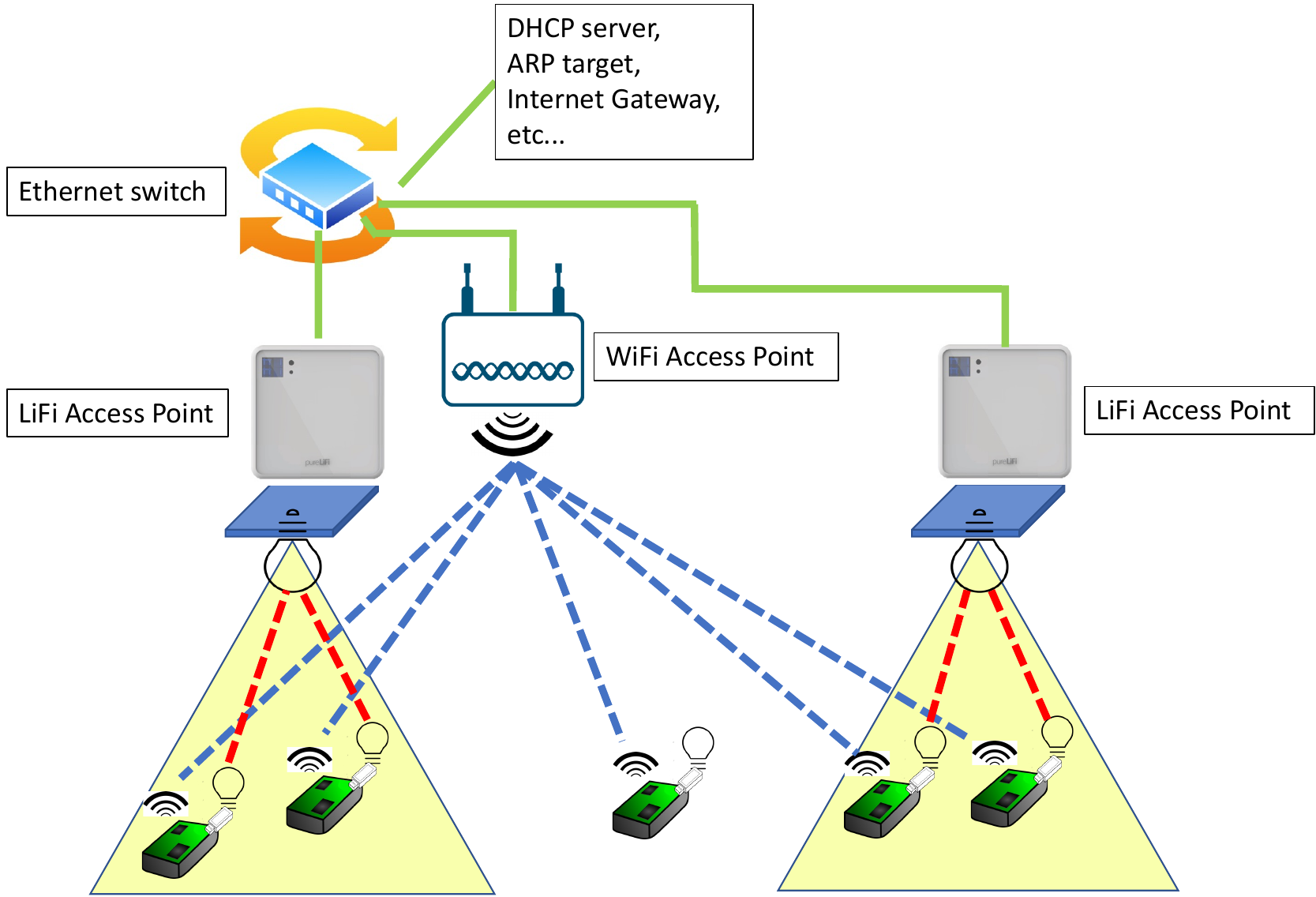}\vspace*{-1mm}
\caption{Hybrid LiFi/WiFi network architecture}
\label{fig:network-architecture}
\end{center}\vspace*{-9mm}
\end{figure}

A typical hybrid LiFi/WiFi network layout is represented in Figure~\ref{fig:network-architecture}. LiFi and WiFi APs are present in the network to serve mobile devices. One switch (or more, in practical scenarios) is present to connect the LiFi and WiFi branches, allowing any node to communicate through either of its two physical network interfaces (each using its own wireless technology) with any other node, even if the destination is temporarily using a different type of interface. A server, shown at the top of the figure, which provides basic services such as gateway and DHCP functions, is also present. Typical deployment scenarios for this kind of network are, amongst others, office, industrial, hospital scenarios \cite{Wu2021,Jarchlo2022}.

In the considered network architecture, some mechanism is in place to monitor the status of the link of both interfaces of each mobile device with their respective AP. The mechanism is tasked  with activating a vertical handover upon the verification of certain conditions. This function can be executed at different layers of the protocol stack. For Linux-based systems, there are two official tools operating at the MAC layer that can be used to effectively enable vertical handovers using COTS: the Bonding Driver \cite{linuxbonding} and the Libteam library \cite{libteam}. Both tools can be configured to use either of two means to monitor the status of the link on a network interface (wireless, in our case).

The first means is named \textquotedblleft ARP monitoring\textquotedblright\ (after Address Resolution Protocol, [11]) in the Bonding Driver terminology and \texttt{arp\_ping} in the Libteam terminology. ARP monitoring is based on the transmission of dedicated ARP messages. Each mobile device periodically sends broadcast ARP requests through the network. The ARP request includes the indication of one or more network devices, called \textquotedblleft ARP targets,\textquotedblright\ whose IP address is specified by the \texttt{arp\_ip\_target} (for the Bonding Driver) or \texttt{target\_host} (for Libteam) configuration parameters. The ARP targets are entitled to reply to each ARP request with a unicast ARP reply, sent back to the sender of the ARP request. In the configuration shown in Figure~\ref{fig:network-architecture}, the ARP target could be set to be the central server.

The second means is named \textquotedblleft MII monitoring\textquotedblright\ (after Media Independent Interface,~\cite{IEEESTD802.3-2018}). MII monitoring can use either a \texttt{ETHTOOL\_GLINK} ioctl call of the \texttt{ethtool} utility \cite{ethtool}, if properly supported by the network interface card (NIC) driver, or the \texttt{netif\_carrier\_ok()} Linux kernel function which, if supported, is a preferable option. The NIC driver replies to these calls with the status of the link.

For connectivity losses due to impairments of the physical (either VLC or RF) channel, MII monitoring has a much slower response time \cite{Pescosolido2022,Ancillotti2025}. Furthermore, we have verified that
the Libteam tool generally performs better than the Bonding Driver, while also conveying greater reliability and  configuration flexibility. These findings will be reported in \cite{Ancillotti2025}. For this reason, in this work, we adopt the Libteam library, configured to use ARP monitoring, as the baseline solution for vertical handovers.\vspace*{-1mm}

\subsection{Testbed description\label{subsec:testbed-description}\vspace*{-1mm}}
\begin{figure}[t]
\begin{center}
\includegraphics[width=0.8\columnwidth]{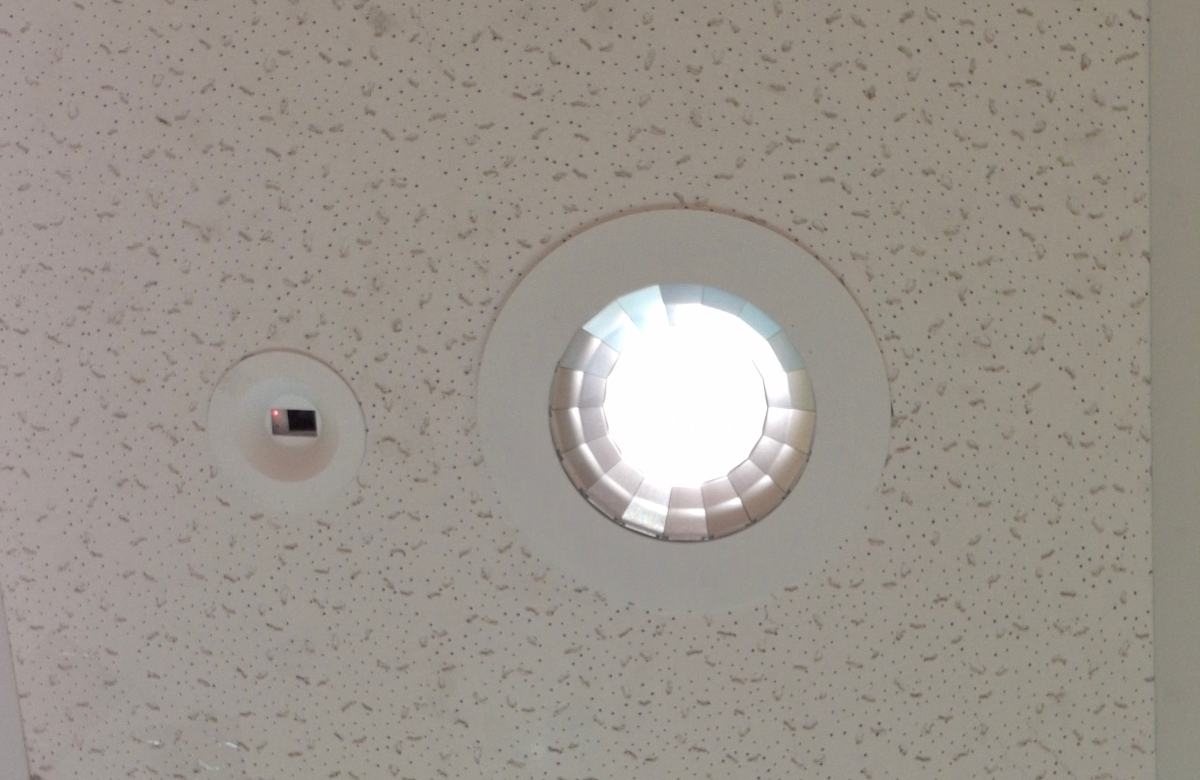}\vspace*{-2mm}
\caption{LED lamp and IR sensor on the ceiling}
\label{fig:LED-lamp}
\end{center}\vspace*{-5mm}
\end{figure}

Our testbed is composed of two PureLiFi \cite{PureLifi} XC Access Points mounted on the ceiling of our lab, at 3.15~m height. Each AP is equipped with a 20W 4000 K Lucicup II LED Lamp by Lucibel, with a maximum luminous power of 1930~lm, which transmits downlink VLC signals, and an infrared sensor receiving uplink IR signals from the wireless devices. Both items are shown in Figure~\ref{fig:LED-lamp}. For the purposes of this work, only
one of the two APs has been used.

\begin{figure}[t]
\begin{center}
\includegraphics[width=0.8\columnwidth]{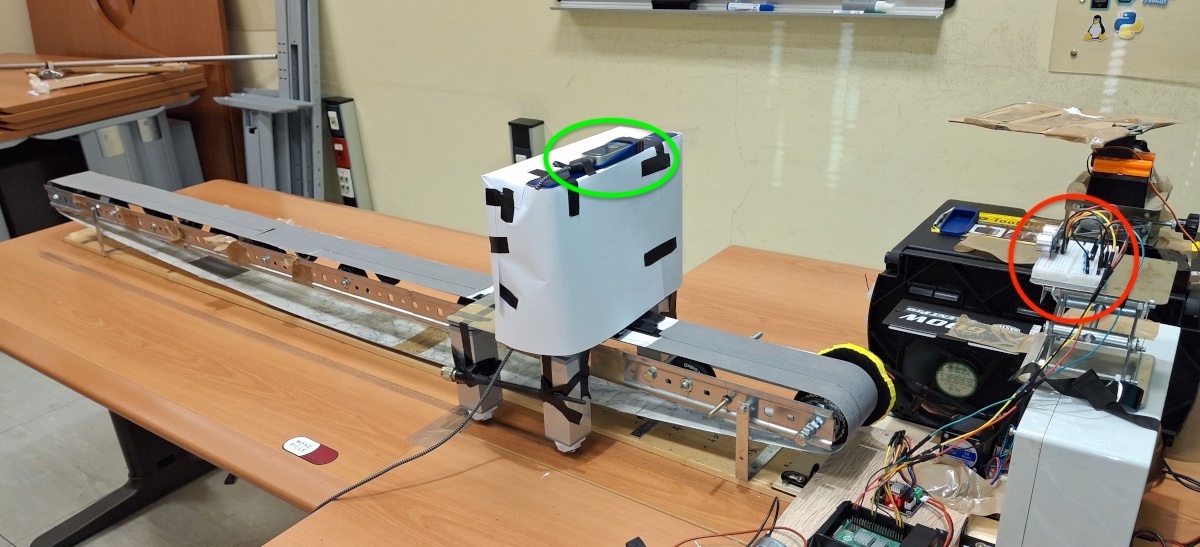}\vspace*{-2mm}
\caption{Conveyor Belt}
\label{fig:conveyor-belt}
\end{center}\vspace*{-8mm}
\end{figure}

The wireless devices consist of six ADJ 270-00108 PC sticks, equipped with an Intel Atom Z8350 processor, 2 GB RAM and 32 GB eMMC hard disk, 802.11 a/b/g/n/ac WiFi card, Bluetooth 4.0, one USB 2.0 port, one USB 3.0 port, one HDMI port. In the PC sticks, we installed the Ubuntu 22.04 Operating System (OS). To each PC Stick is attached a PureLiFi LiFi-XC Station USB dongle, which transmits uplink IR signals through an IR LED and receives VLC signals from the AP Lamp through a photodiode receiver.

A Gateworks Ventana GW5300 single board computer (NXP\texttrademark~i.MX6 800MHz Dual Core ARM\textregistered~Cortex\texttrademark~-A9 processor, 1GByte RAM), acts as a WiFi AP. 

A conveyor belt, shown in Figure~\ref{fig:conveyor-belt}, activated by a motor, is used to enable device mobility. One of the LiFi dongles, circled in green in Figure~\ref{fig:conveyor-belt}, is attached to the conveyor belt. The other devices are placed on the table in the same attocell. Their role in the experiments reported in this work is to add realism in terms of presence of signalling traffic from multiple users, mostly due to the ARP messages used by the Libteam mechanism. They do not generate, however, data traffic. 

When the conveyor belt is activated, an ultrasound ranging device, circled in red in Figure~\ref{fig:conveyor-belt}, is used to precisely map the position of the LiFi dongle to specific time instants.

Listing~\ref{lst:json-config} shows the JSON configuration file for the Libteam daemon used in our experiments. The interested reader is encouraged to check the documentation for the Libteam configuration \cite{libteam-conf}. The value assigned to the different parameters have been selected on the basis of an extensive experimental analysis which we will present in~\cite{Ancillotti2025}.\vspace{-2mm}

\begin{listing}\vspace{-2mm}
\begin{lstlisting}
{
    "device":   "team0",
    "notify_peers": {
        "count": 4,
        "interval": 50
    },
    "runner":   {
        "name": "activebackup",
        "hwaddr_policy": "by_active"
    },
    "ports":    {
        "wlp1s0": {
            "prio": 1,
            "link_watch":   {
                "name": "arp_ping",
                "interval": 100,
                "missed_max": 2,
                "target_host": "192.168.1.2",
                "validate_active": false,
                "validate_inactive": false,
                "send_always": true
            }
        },
        "wlx70b3d5958671": {
            "prio": 2,
            "link_watch":   {
                "name": "arp_ping",
                "interval": 100,
                "missed_max": 2,
                "target_host": "192.168.1.2",
                "validate_active": false,
                "validate_inactive": false,
                "send_always": true
            }
        }
    }
}
\end{lstlisting}\vspace{-2mm}
\caption{Libteam JSON configuration file}
\label{lst:json-config}
\end{listing}\vspace{-4mm}

\section{Problem statement}\label{sec:problem-statement}\vspace{-1mm}
Data-link layer solutions, such as the Bonding Driver or Libteam, are able to perform a vertical handover in a relatively short time, but the trigger for it is essentially a connectivity loss. In situations with QoS requirements, say in terms of achievable goodput, QoS-referred out-of-service intervals of duration on the order of several seconds occur. In the following, we illustrate the problem by experimental results obtained in the baseline configuration described above.

\begin{figure}[!t]
\centering
\subfloat[Goodput in static positions]{\includegraphics[width=0.9\linewidth]{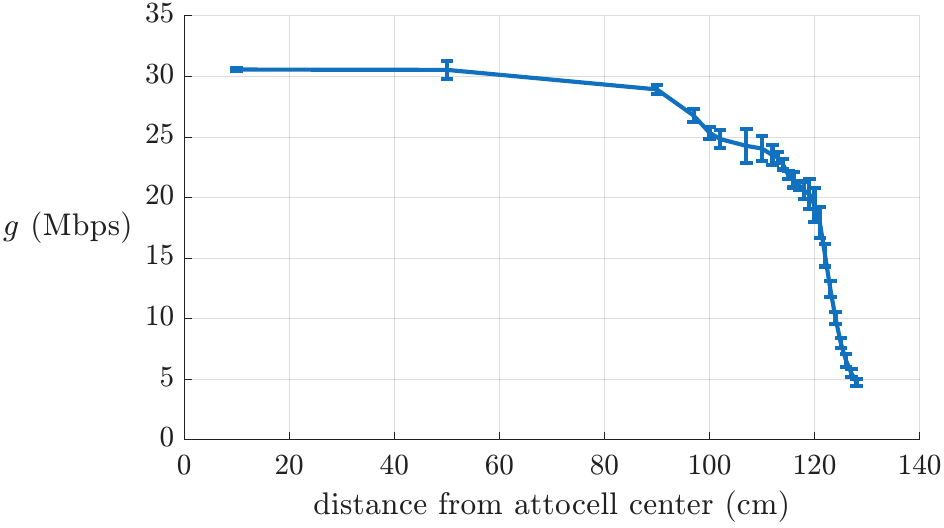}\label{fig:throughputVSdistanceStatic}}\vspace*{-1.5mm}
\hfil
\subfloat[Goodput samples at a speed equal to 0.15 m/s]{\includegraphics[width=0.9\linewidth]{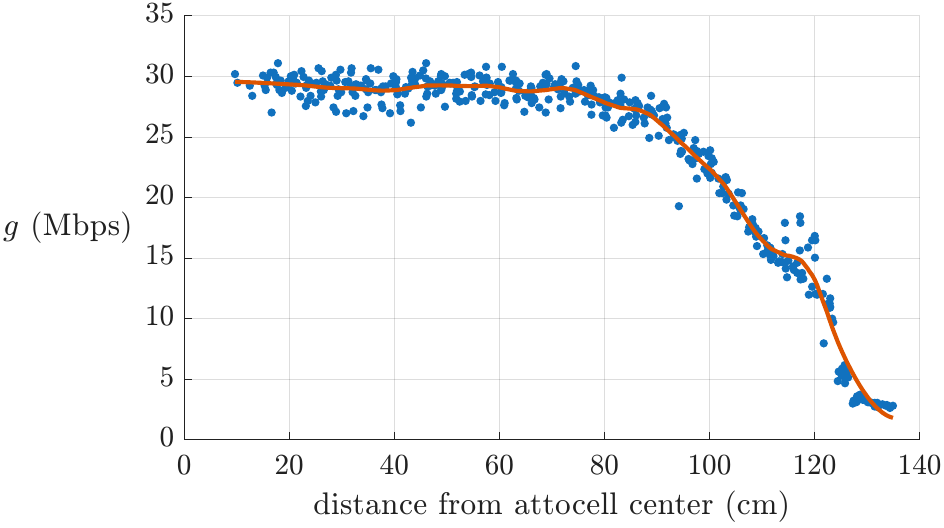}%
\label{fig:throughputVSdistanceDynamicPointCloud}}\vspace*{-1.5mm}
\hfil
\subfloat[Goodput averages at different speeds]{\includegraphics[width=0.9\linewidth]{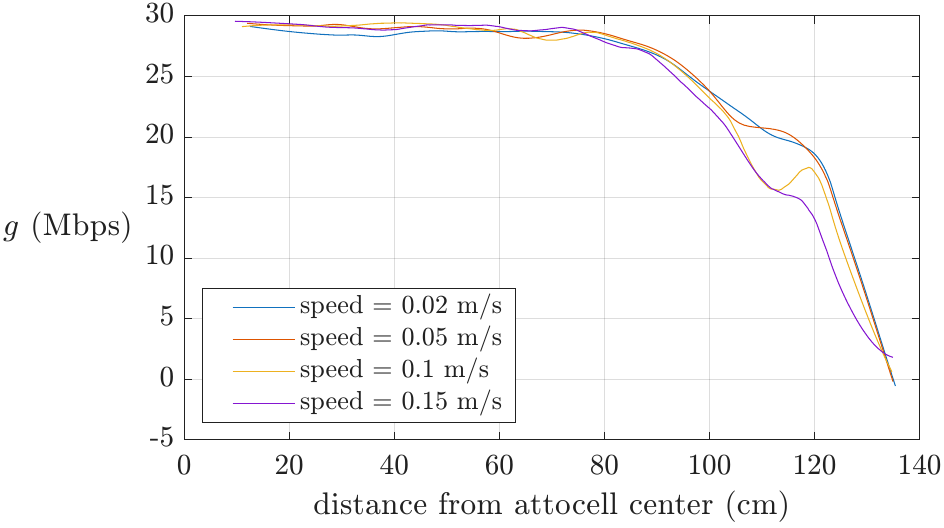}%
\label{fig:throughputVSdistanceDynamic}}\vspace*{-1mm}
\caption{Goodput measured at different distances under various speed conditions}
\label{fig:throughputVSdistance}\vspace*{-7mm}
\end{figure}

The coverage region of a LiFi attocell is a conical 3D region with apex at the LED lamp. A different coverage radius applies to each plane placed at a given height. Without loss of generality, in the following, we refer to the testbed deployment described in Section~\ref{sec:network-model}. With this deployment, LiFi connectivity is reliable up to (almost) 135~cm from the attocell center. However, the performance is far from being uniform in this region. Figure~\ref{fig:throughputVSdistance} shows the average downlink UDP goodput, $g$, achieved by a mobile device using the LiFi interface at different distances from the attocell center, ranging from 10~cm to 135~cm. UDP packets with a 1470 bytes payload were injected at a (payload) rate of 40~Mbps. The injection rate is larger than the maximum UDP goodput that can be supported by the LiFi interface of our devices, even in the absence of packet errors, In this way, the results reflect the maximum practical goodput achievable in each device position and moving speed.

Subfigure~\ref{fig:throughputVSdistanceStatic} shows the results obtained in a static configuration: the device has been kept at each position for 60 seconds, and goodput samples were acquired at intervals of one second. The experiment has been replicated 10 times, for an overall number of 600 goodput samples for each distance. The plot shows the average goodput and its standard deviation, computed, for each distance, over the 600 samples.

The point cloud in Subfigure~\ref{fig:throughputVSdistanceDynamicPointCloud} shows the goodput samples in a dynamic scenario: the device moves at a speed of 0.15 m/sec starting from 10~cm off the attocell center along a radial path. The represented goodput samples have been collected every 200~ms during the device motion, in 10 experiment replicas. Using the ultrasound ranging device, we collected the position at each goodput sampling instant, which allows us to represent the results as a function of the spatial coordinate. The red solid line on top of the point cloud has been obtained by a weighted least squares local regression on the points in the cloud, using a 0.2 span coefficient. It can be considered as a proxy for the average value for each distance.

The plots in Subfigure~\ref{fig:throughputVSdistanceDynamic} show the same type of average goodput, obtained in dynamic conditions, as the solid line in Subfigure~\ref{fig:throughputVSdistanceDynamicPointCloud}, referred to the same type of experiment, but executed with different device motion speeds.

All the plots in Figure~~\ref{fig:throughputVSdistance} have a common feature: three regions can be identified. The first region extends from the cell center to roughly 90~cm from it. In this region, the average goodput is relatively stable, with value slightly below 30~Mbps, which slightly decreases with increasing distance. This holds in all speed conditions. From 90~cm to 120~cm the goodput decreases more sharply with distance. However, it is important to note that, in this region, the slope of the curve depends on the speed. The faster the motion, the steeper the slope. Beyond 120~cm, the slope further increases, with connectivity being lost at around 135~cm, or slightly more in some cases.

The difference in the slopes of the curves in the region beyond 90~cm depends on the fact that the protocol stack of the LiFi components (AP and USB dongles) includes an adaptation mechanism for the channel conditions. The hardware producer has not provided the details of such implementation, which is not standard-compliant, but we have been able to verify it by experience. As the communication performance deteriorates over time, the MAC reacts by lowering the injected payload data rate, likely adding redundancy or adopting a more robust modulation scheme. At higher speeds, this mechanism has less time to react to the quickly varying conditions, and the communication tends to be more prone to packet reception errors.

\begin{figure}[t]
\begin{center}
\includegraphics[width=0.9\columnwidth]{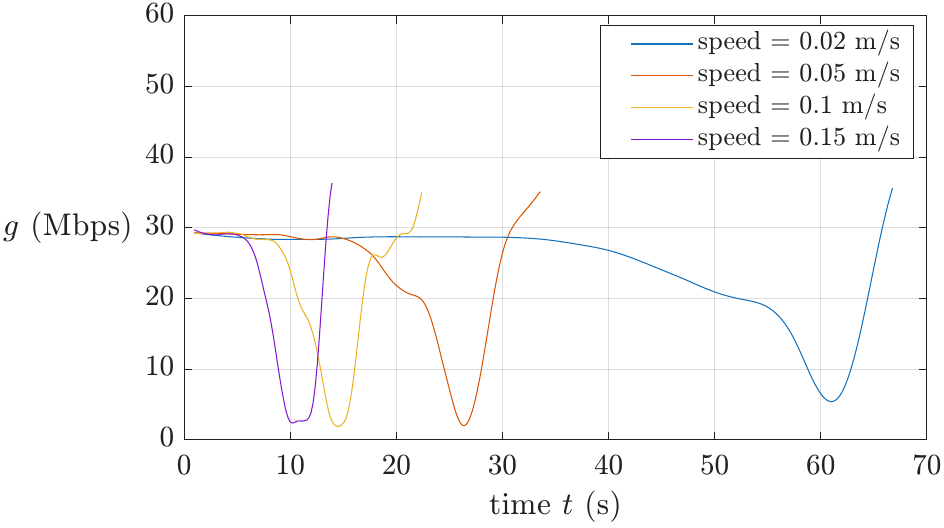}\vspace*{-2mm}
\caption{Goodput averages at different speeds}
\label{dynamicThroughputVStimeLOESS}
\end{center}\vspace*{-8mm}
\end{figure}

Figure~\ref{dynamicThroughputVStimeLOESS} shows results from the same experimental traces as those used for Subfigure~\ref{fig:throughputVSdistanceDynamic}, still using a weighted least squares local regression on the samples, but plotted against a temporal axis. The intervention of the Libteam-enabled vertical handover after the loss of LiFi connectivity is evident in the rightmost part of each plot, with the abrupt increase in goodput due to the switch to the WiFi interface.

Solutions such as the Bonding Driver or Libteam are able to perform a vertical handover in a relatively short time, but the trigger for it is essentially a connectivity loss. In situations in which there are QoS requirements in terms of a minimum guaranteed achievable goodput, QoS-referred out-of-service intervals of duration on the order of several seconds may occur, depending on the specific requirement, as goodput falls below the desired value at a distance from the center much smaller than the attocell radius, and well before the handover mechanism kicks in.\vspace*{-1.7mm}

\section{Proposed mitigation strategies}\vspace*{-1.3mm}

To mitigate the impact on QoS of the performance decay in the proximity of the attocell border, we start by considering that setting a threshold on the running goodput itself, to trigger a vertical handover, may not be practical, because the goodput displays relevant temporally local variations and would entail the use of an over-conservative threshold to avoid unnecessary frequent vertical handovers. Besides this, in a practical multi-user scenario (not considered in this work), the per-user goodput would depend on the number of concurrent users and on their packet flow rate, as the overall communication resources of the LiFi attocell are shared among the users with active packet flows \cite{Pescosolido2021}. Ideally, we aim at setting the threshold on a metric which evolves in a relatively smooth way over time, and is independent of the presence of other users. 

Let us start by considering, as a candidate metric, the signal strength. Each network interface provides information on the received signal power level. This information can be obtained by using standard tools. For instance, the \texttt{\footnotesize iwtool} library command

\begin{figure}[t]
\begin{center}
\includegraphics[width=0.9\columnwidth]{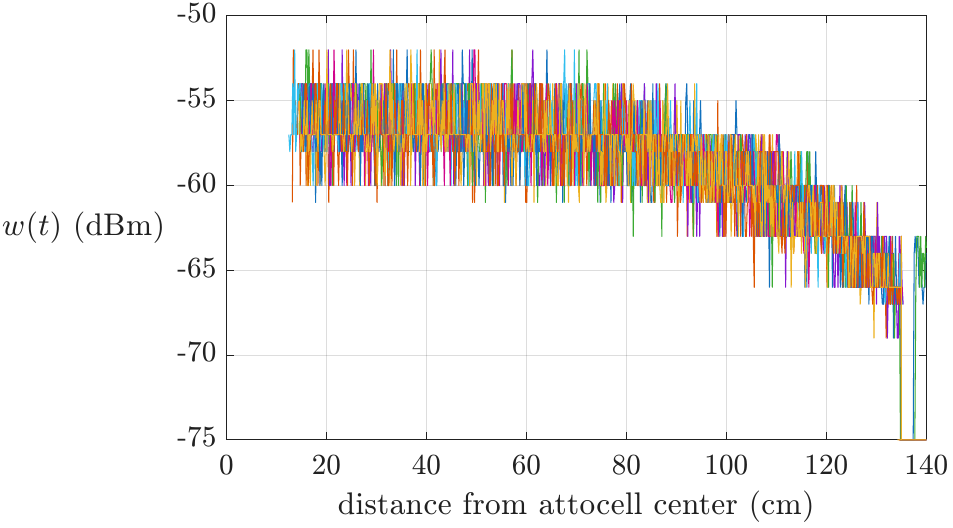}
\caption{Signal power samples acquired with the device moving at 0.02~m/s}
\label{fig:signalVSdistanceDynamic}
\end{center}\vspace*{-5mm}
\end{figure}

\hspace{1em}\texttt{\footnotesize {iw dev [intf] link}},

\noindent where \texttt{\footnotesize [intf]} is the interface name seen by the OS, returns information about the current status of the connection on that interface. Among other information, the received signal power is provided, measured in dBm. The method described in the following is based on regularly querying the interface to obtain the signal power level. We denote $w(t)$ the signal power level, in dBm, reported by the network interface at a given time $t$, and $\delta t$ the polling interval. Figure~\ref{fig:signalVSdistanceDynamic} shows the signal readings sequences taken in the 10 replicas, superimposed with different colors, of the experiment with the device moving at speed 0.02~m/s already described in the previous section. The signal power level readings were acquired every $\delta t = 100 \text{ms}$.

\begin{figure}[t]
\centering
\subfloat[Signal power EWMA @speed = 0.02 m/s]{\includegraphics[width=.5\linewidth]{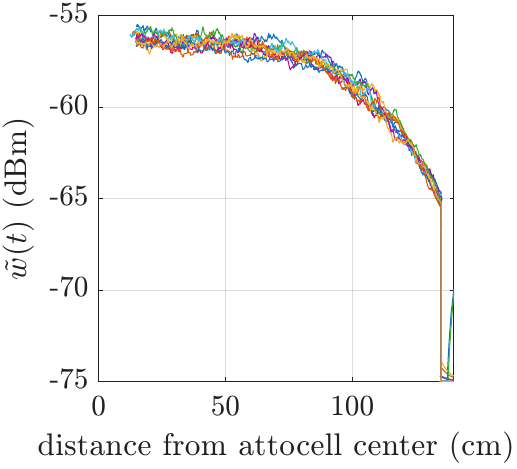}\label{fig:EWMAVSdistanceDynamic0.02}}
\subfloat[Signal power EWMA @speed = 0.05 m/s]{\includegraphics[width=.5\linewidth]{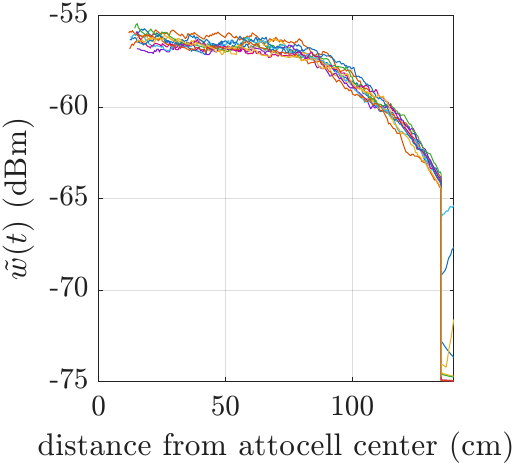}\label{fig:EWMAVSdistanceDynamic0.05}}
\hfil
\subfloat[Signal power EWMA @speed = 0.1 m/s]{\includegraphics[width=.5\linewidth]{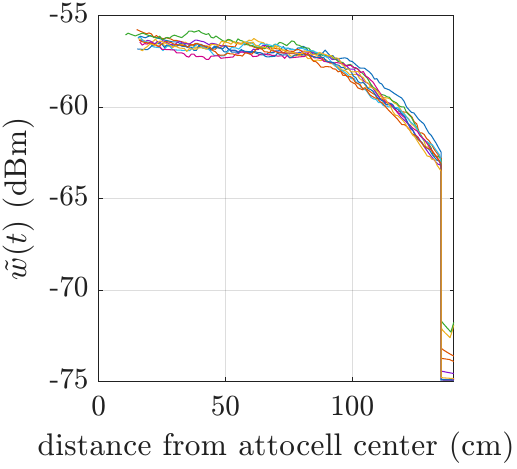}\label{fig:EWMAVSdistanceDynamic0.1}}
\subfloat[Signal power EWMA @speed = 0.15 m/s]{\includegraphics[width=.5\linewidth]{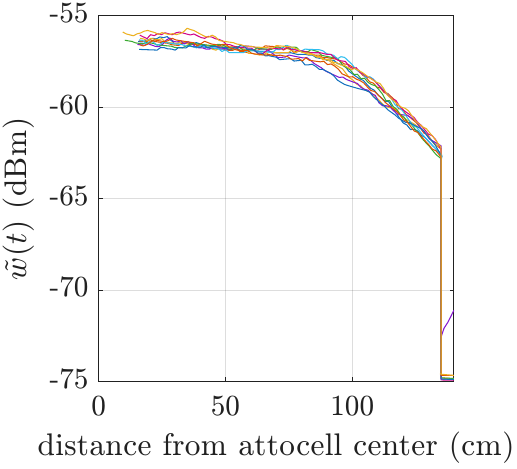}\label{fig:EWMAVSdistanceDynamic0.15}}
\caption{Signal power EWMA, acquired with the device moving at different speeds.}
\label{fig:EWMAVSdistanceDynamic}\vspace*{-5mm}
\end{figure}

This metric, similarly to the instantaneous goodput, presents relatively wide oscillations even across close-by sampling instants. Therefore, intuitively, directly using it would entail over-conservative threshold settings, in order to avoid frequent undesired handovers.

A more suitable metric would be a smoothed version of the signal power sample sequence. In the following, we consider the exponentially weighted moving average (EWMA) of the signal power level readings, defined as\vspace{-2mm}
\begin{equation}
\tilde{w}(t)=(1-\alpha)\tilde{w}(t-\delta t)+\alpha w(t),\vspace{-2mm}
\end{equation}
where $\alpha$ is the smoothing coefficient.

Figure~\ref{fig:EWMAVSdistanceDynamic} shows the signal power level EWMA of the 10 replicas of each of the experiments with the device moving at different speeds. Each subfigure reports the results obtained with a specific speed. The EWMA was computed by setting $\alpha=0.05$. The value for $\alpha$ was selected through experience. An analytical optimisation of this parameter is outside the scope of this work and will be considered in our future work.

It can be seen that this metric provides a much narrower spread across adjacent instants and also different experiment realizations, suggesting that it can be effectively used to trigger vertical handovers in a reliable way.

Observing the plots, we can see that the bundle of curves of the signal EWMA as a function of distance tends to be wider at lower speeds. This is a direct consequence of the fact that, at lower speeds, the number of temporal samples corresponding to a given distance range is larger. For the same reason, the higher the speed, the larger the distances of past signal samples that are incorporated in the EWMA from the present position.

A shortcoming of setting a threshold on the signal power EWMA to trigger vertical handovers is the following: recalling the discussion about the dependence of the goodput degradation in the region between 90 and 120~cm, see Subfigure~\ref{fig:throughputVSdistanceDynamic}, the behavior of the EWMA as a function of speed is of no help. In fact, given a certain threshold, the inherent low-pass and causal nature of the EWMA tends to postpone the vertical handover with increasing value of the speed, whereas, as seen before, the goodput degrades faster. This suggests that, for a given scenario, in which the maximum speed at which nodes move is known, the threshold should be set in a conservative way, i.e., using the mapping between signal power EWMA and goodput for the maximum foreseeable speed.

To mitigate this problem, we consider an additional metric which we propose to use: the running estimate of the packet reception failure percentage. This is another metric which is typically made available by network interfaces. It exploits the Cyclic Redundancy Check (CRC) bits appended to each packet. We obtain it with the command

\hspace{1em}\texttt{\footnotesize {ethtool -S \$[intf] | grep -e phy\_rx\_crc\_err -e rx\_packets}},

\noindent where \texttt{\footnotesize [intf]} is the interface name defined in the OS. The command returns the ratio of packets that have failed the CRC test to the overall number of packets received in a given interval, i.e., it is a number in the range between 0 and 1.

\begin{figure}[t]
\centering
\subfloat[Example of CRC-based packet failure ratio and its EWMA, @speed = 0.02 m/s]{\includegraphics[width=.9\linewidth]{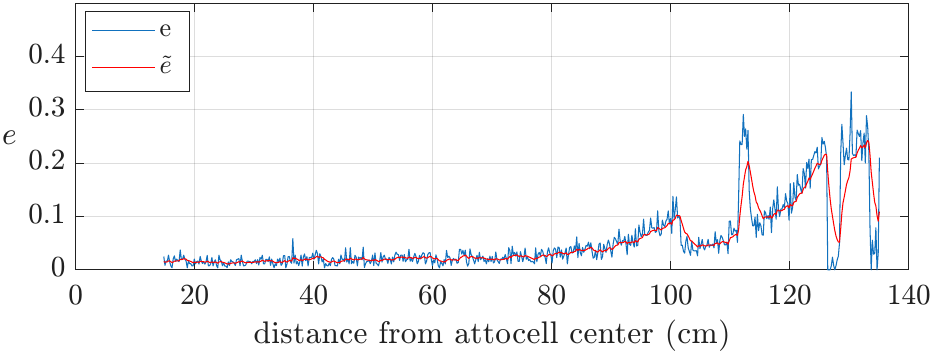}\label{fig:CRCexample1}}\vspace*{-1.5mm}
\hfil
\subfloat[Example of CRC-based packet failure ratio and its EWMA, @speed = 0.15 m/s]{\includegraphics[width=.9\linewidth]{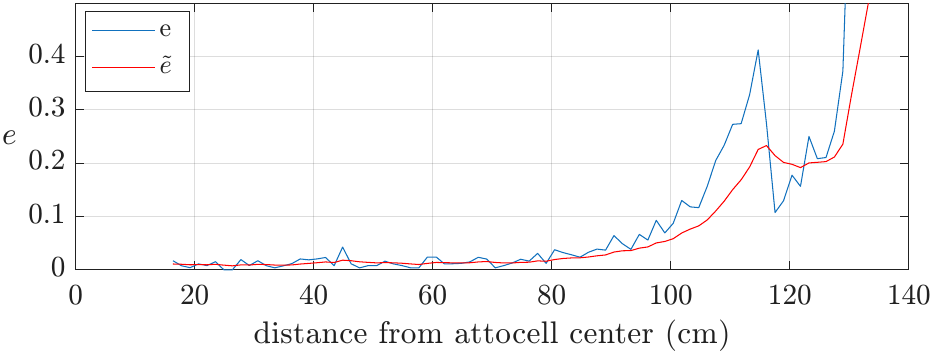}\label{fig:CRCexample2}}
\caption{CRC-based packet failure ratio and associated EWMA}
\label{fig:CRCexample}\vspace*{-5mm}
\end{figure}

In the following, we denote the CRC-based packet failure ratio, obtained at time $t$, as $e\left(t\right)$. Figure~\ref{fig:CRCexample} shows an example of the evolution of the CRC-based packet failure ratio (blue plots) as a function of distance, obtained at speed equal to 0.02~m/s (Subfigure~\ref{fig:CRCexample1}), and 0.15~m/s (Subfigure~\ref{fig:CRCexample2}), with samples taken every $\delta t = 100~\text{ms}$. In red, their associated exponentially weighted moving averages computed with $\alpha = 0.05$.

It is worth noticing that, in the example with speed 0.02~m/s (Subfigure~\ref{fig:CRCexample1}), in the critical region beyond 90~cm the metric has several ups and downs. This reflects the intervention of the mechanism, already mentioned in Section~\ref{sec:problem-statement}, which adapts the robustness of the transmission to the communication performance. At higher speeds, see Subfigure~\ref{fig:CRCexample2}, the mechanism is less effective, leading to the performance differences, in terms of goodput decay, shown in Subfigure~\ref{fig:throughputVSdistanceDynamic}.

The presence of adaptive mechanisms of this type in the network interface drivers, suggests that the handover method based on CRC-based packet failure ratio should in any case incorporate a confirmation check on the signal power level, which should be below a given threshold to confirm the execution of the handover. Otherwise, the handover mechanism would conflict with the robustness adaptation mechanism, preventing it from intervening. The handover would likely be activated, depending on the threshold set for it, when there is (still) no need to do it.

Besides this, as for the case of signal power level readings, to avoid the effect of local fluctuations, we consider the exponentially weighted moving average of the CRC-based packet failure ratio readings. Sample traces of the CRC-based packet failure ratio EWMA are the red plots in Figure \ref{fig:CRCexample}.

While the CRC-based packet failure ratio EWMA is subject, qualitatively, to the same problem as the signal power level EWMA, in terms of incorporating samples from the past, we will show in the next section that using it, in combination with the signal level, provides an improvement in the vertical-handover performance in terms of QoS outage duration.\vspace{-2mm}

\section{Performance evaluation}\label{sec:perf-eval}\vspace{-1mm}
We have evaluated the proposed vertical handover methods in several configurations at different speeds. As discussed in Section~\ref{sec:problem-statement}, when taking into account different speeds, there is no unique correspondence between signal power levels and goodput, even in terms of average values. Therefore, setting the thresholds in order to pursue a QoS constraint is not trivial. For the purposes of this work, we have selected 20 Mbps as the QoS requirement for the goodput. Then, several combinations of thresholds have been selected, based on our experience with the measurements, with the objective of illustrating the performance trends of the proposed techniques.

Let $\lambda_{\tilde{w}}$ and $\lambda_{\tilde{e}}$ be the thresholds used by the two proposed vertical handover methods. With the method based on the signal power level EWMA, we have considered the thresholds $\lambda_{\tilde{w}} = -62 \text{dBm}$ and $\lambda_{\tilde{w}} = -64 \text{dBm}$. With the method using the CRC-based packet failure ratio EWMA, we have considered the thresholds $\lambda_{\tilde{e}} = 0.15$ and $\lambda_{\tilde{e}} = 0.20$, each in combination, for the confirmation check, with either of the two values of $\lambda_{\tilde{w}}$.
\begin{figure}[p]
\centering
\subfloat[Performace at speed = 0.02 m/s]{\includegraphics[width=.83\linewidth]{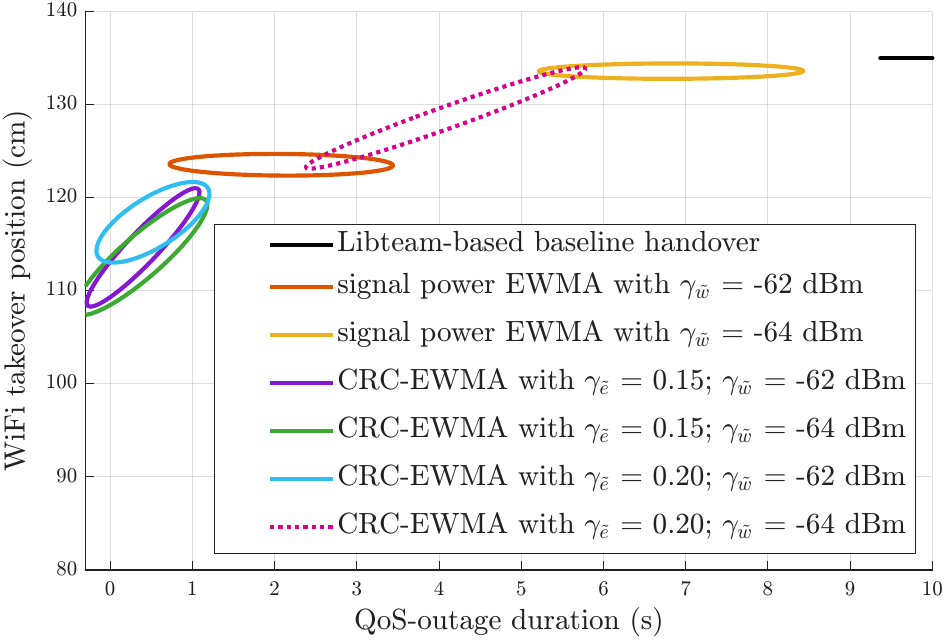}\label{fig:perfEval1}}\vspace*{-1.5mm}
\hfil
\subfloat[Performace  at speed = 0.05 m/s]{\includegraphics[width=.83\linewidth]{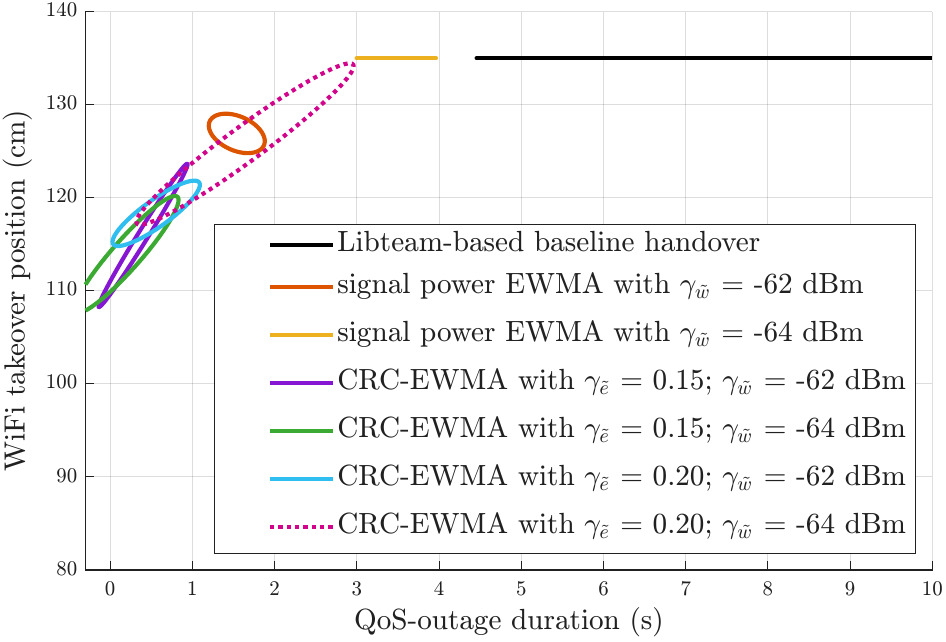}\label{fig:perfEval1}}\vspace*{-1.5mm}
\hfil
\subfloat[Performace  at speed = 0.1 m/s]{\includegraphics[width=.83\linewidth]{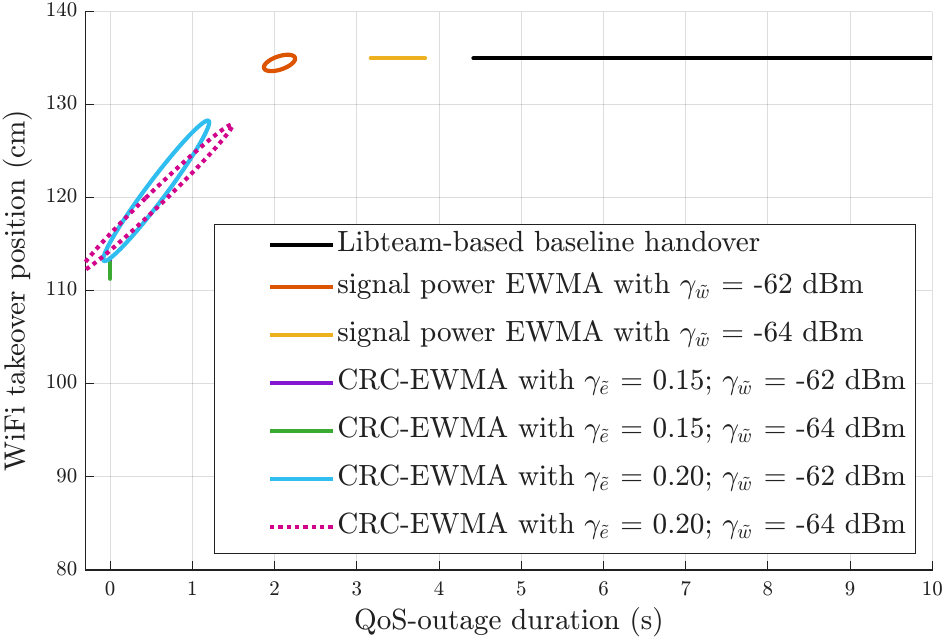}\label{fig:perfEval1}}\vspace*{-1.5mm}
\hfil
\subfloat[Performace  at speed = 0.15 m/s]{\includegraphics[width=.83\linewidth]{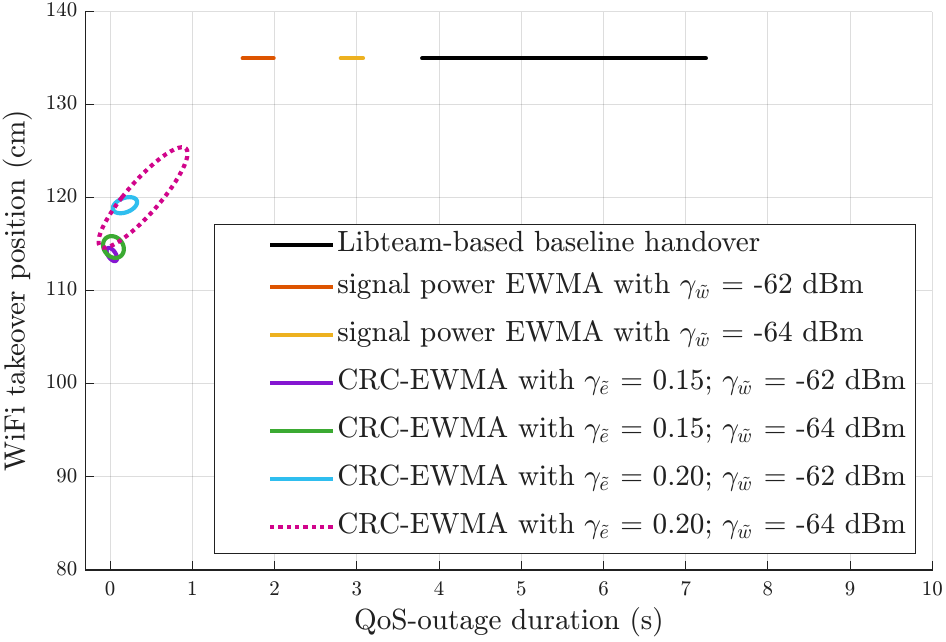}\label{fig:perfEval1}}
\caption{Covariance ellipses of QoS-outage duration and WiFi takeover distance of the considered techniques evaluated for different motion speeds.}
\label{fig:PerfEval}\vspace*{-5mm}
\end{figure}

Figure~\ref{fig:PerfEval} reports the results of our experiments with the proposed handover methods. Each subfigure refers to a different speed of the mobile device, which starts its motion in a radial path from a position located slightly (15~cm) off the attocell center. The horizontal axis reports the QoS outage durations, i.e. the duration of the interval during which $g(t) < 20\text{ Mbps}$. To be more precise, to avoid the effect of local fluctuations of $g(t)$, to compute the QoS-outage duration we have used a time-symmetrical smoothed version of it, obtained with a weighted least squares local regression on $g(t)$, using a 0.2 span coefficient. The vertical axis reports the distance from the attocell center at which the WiFi communication has reached a steady state. Distances above 135~cm (outside the attocell border) are nominally represented by the value of 135~cm. Visualizing this WiFi intervention distance allows evaluating the portion of the nominal attocell coverage area which gets eroded by the intervention of the mechanism.

Performance is presented as covariance ellipses of the 2-dimensional data set (QoS outage duration and WiFi takeover distance) provided, for each experiment, by 10 replicas. We use the covariance elliptic representation, as opposed to box plots, as it captures the trade-off between the desire to reduce outage duration with that of maintaining a relatively large \emph{experimental} LiFi coverage. Because both measured metrics are random quantities, determined by the specific replica of the experiment, this representation allows capturing the performance in terms of the above mentioned trade-off, besides providing information on the mean values obtained (the center of the ellipses) and the reliability of the estimate (the width of the axis). The closer to the upper-left corner a method's performance lies, the better.

As expected, with the baseline Libteam solution, because the handover is triggered by a connectivity loss, the QoS-outages may last for several seconds. The methods based on the signal power level EWMA perform better, reducing the outage duration to the 2–3 second range, depending on which threshold is used. We point out, however, that the outage duration exhibits a strong dependence on the speed of motion.

In contrast, the methods involving the CRC-based packet failure rate have performance much less dependent on the device speed, as the outage duration stays consistently below 1 second.

The best-performing combination, among the considered ones, is the one based on the CRC failure reports, with threshold set at $\lambda_{\tilde{e}} = 0.20$, and confirmation threshold set at $\lambda_{\tilde{w}} = -62~\text{dBm}$. In fact, it succeeds in reducing the average QoS-outage duration to below half a second, while keeping the WiFi takeover distance around 120~cm, irrespective of the speed of motion.\vspace{-3mm}

\section{Conclusion}\label{sec:conclusion}\vspace{-1mm}
This work brings a contribution in the area of hybrid LiFi/WiFi networks built out of commercially available devices, to improve the performance of vertical handovers in terms of QoS maintenance in mobile scenarios. We have identified specific aspects of the problem related to device mobility, such as the dependence of performance degradation, near the attocell border, on the speed of motion, which makes devising effective handover methods a challenging task. Another important feature of our proposed techniques is that they take into account the presence of automatic communication robustness adaptation mechanisms in the network interface drivers, thus avoiding conflicting with them.
The paper opens the way to further research challenges such as the optimisation of the run-time metrics filtering methods (of which the EWMA is a special case), possible alternative ways of combinations of the considered on-line metrics and threshold settings, and the extension of the proposed methods to the multi-user case.\vspace{-1mm}

\section*{Acknowledgment}\vspace{-1mm}
This work was partially funded by the European Union under the Italian National Recovery and Resilience Plan (NRRP) of NextGenerationEU partnership on Telecommunications of the Future (program PE00000001 - RESTART).\vspace{-1mm}

\bibliographystyle{IEEEtran}

\end{document}